\documentclass[cits]{PoS}

\def\la{\mathrel{\hbox{\rlap{\hbox{\lower4pt\hbox{$\sim$}}}\hbox{$<$}}}}
\def\ga{\mathrel{\hbox{\rlap{\hbox{\lower4pt\hbox{$\sim$}}}\hbox{$>$}}}}

\title{Supergiant Fast X-ray Transients observed by INTEGRAL}

\ShortTitle{SFXTs observed by INTEGRAL}

\author{\speaker{Sergei A. Grebenev}\\ 
        Space Research Institute (IKI), Russian Academy of Sciences\\
        Profsoyuznaya 84/32, 117997 Moscow, Russia\\
        E-mail: \email{sergei@hea.iki.rssi.ru}}

\abstract{We review X-ray properties of the Supergiant Fast
X-ray Transients following from their observations with INTEGRAL
and show that a compact object in these systems is a neutron
star with strong magnetic field accreting from the stellar wind
of a donor star. We show that presence of a centrifugal barrier
at the magnetospheric boundary of the neutron star may be a key
to understanding of abrupt short X-ray outbursts of these
transients and long intervals of their quiescence.}

\FullConference{The Extreme sky: Sampling the Universe above 10 keV\\
		October 13-17, 2009\\
		Otranto (Lecce), Italy}

\begin{document}

\section{Introduction}
Only a few of $\sim 50$ high mass X-ray binaries (HMXBs) known
before the launch of INTEGRAL had supergiants of early (O-B)
spectral types as optical counterparts (most of HMXBs contained
Be-stars). In spite of different origin of compact objects
(e.g., a black hole in \mbox{Cyg X-1} or neutron stars in Vela
X-1 and 4U\,1700-37) all these HMXBs were bright
quasi-persistent X-ray sources powered completely or in part by
accretion from the stellar wind of the supergiant. The
observed amount of such sources, small in comparison with
theoretical expectations, was explained by inhibition of steady
accretion in a numerous `dark' population of the sources due to
a centrifugal barrier at the magnetospheric boundary of the
neutron star -- the so-called ``propeller effect''
\cite{illarionov75}.

The situation became more complex with INTEGRAL that (thanks to
its high sensitivity in hard X-rays, large fields of view of
main telescopes and long uninterrupted observations of the
Galactic disk and Galactic center regions \cite{cw03})
discovered two completely new groups of supergiant X-ray
binaries (SXBs): 1). ``strongly absorbed sources'', enshrouded
in the very dense opaque wind and thus invisible in the standard
X-ray band, and 2). ``supergiant fast X-ray transients''
(SFXTs), flaring up in X-rays for rather a short (one day or
less) time (see for review \cite{negueruela06,sguera06,chaty08}). The
latter sources and actual mechanisms of their outbursts are the
subject of our report.

\section{Observations}
Since the discovery of first SFXTs \cite{sunyaev03b,smith04,intzand04,gs05} 
the amount of known sources of this type increased significantly.
Table 1 lists several broadly discussed SFXTs and their key
parameters: a companion type, spin $P_s$ and orbital $P_b$
periods, a distance $d$. Note that already the detection of
coherent pulsations in some of the sources implies that
their compact object is a neutron star with strong magnetic
field, while the orbital periods $P_b\sim 3.5\!$ -- $\!30$ days imply that
SFXTs could be bright
\begin{table}[h] 
\vspace{-2mm} 
\caption{Known SFXTs and their most important parameters}
\label{table}
\centering
\begin{tabular}{lccccl}
\\ \hline
\multicolumn{1}{c}{Source}&Companion&$P_{s}$&$P_{b}$&$d$\\
\multicolumn{1}{c}{name}   &      type     &s&d&kpc\\ \hline
IGR\,J08408-4503 &O8.5\,Ib(f)&        &          &3  \\
IGR\,J11215-5952 &B0.7\,Ia   &186.8   &165.      &6.2\\
IGR\,J16465-4507 &B0.5\,I/O9.5\,Ia&228.    &        & 9.4\\
IGR\,J16479-4514 & O8.5\,I/O9.5\,Iab&        &  3.32      & 4.9\\
XTE\,J1739-302   &O8\,Iab(f) &        &    12.87 &2.7\\
AX\,J1749.1-2733 &           &131.9    &185.5     &$>8$ \\
IGR\,J17544-2619 &O9\,Ib     &        &  4.93   &3.6\\
SAX\,J1818.6-1703&B0.5-1\,Iab  &        & 30.0  &2.1\\
AX\,J1841.0-0536 &B0\,I/B1\,Ib&4.74   &         &6.9 \\
AX\,J1845.0-0433 &O9.5\,I    &        &          & 3.6   \\
IGR\,J18462-0223 &           &        &          &$\sim6$ \\
IGR\,J18483-0311 &B0-1\,Iab  &21.05   & 18.55    &2.8\\ \hline 
\end{tabular}
\end{table}  
\begin{figure}[t]
\hspace{-1mm}\includegraphics[width=1.02\textwidth]{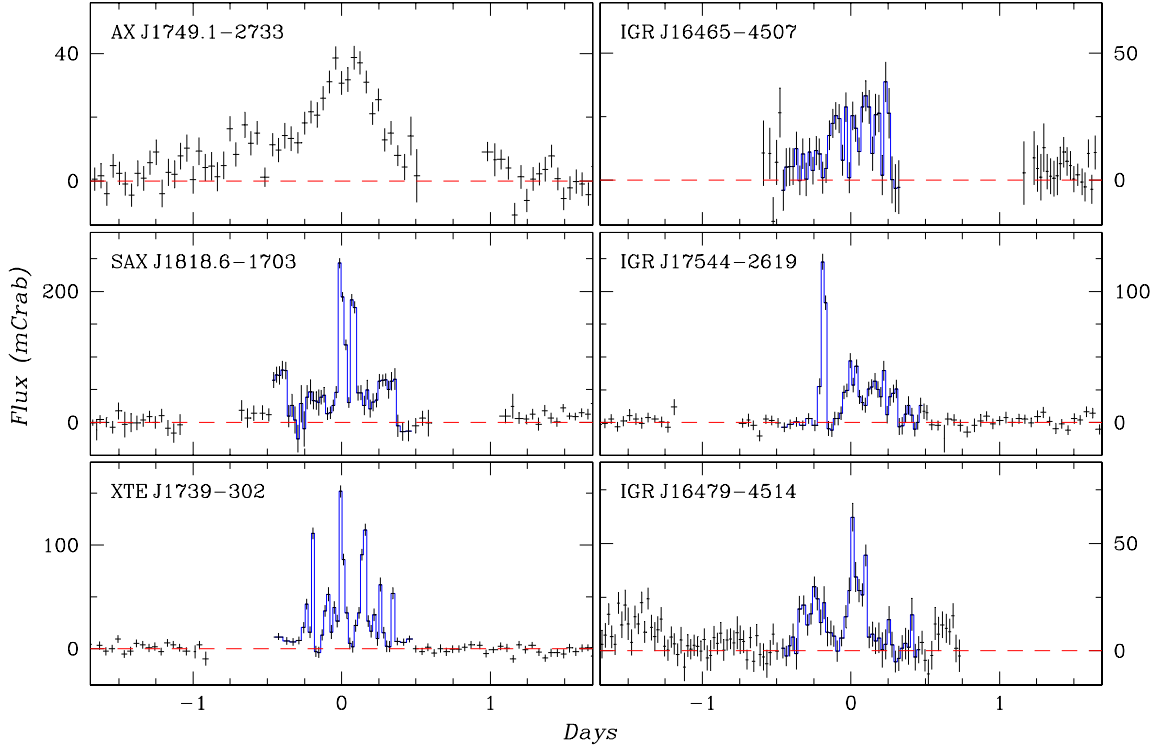}

\vspace{-10mm} 
\caption{IBIS/ISGRI 20-60 keV light curves measured in 2003-2004 
  during first detected outbursts of 6 most reliable SFXTs (epoch ``0'' 
  corresponds to UT
  2003,  
  August 10.4505 and   
         27.0495   
         for IGR\,J16479-4514 and XTE\,J1739-302, 
  September 9.4695, 
            9.5895 and  
           17.2695  
         for SAX\,J1818.6-1703, AX\,J1749.1-2733 and IGR\,J17544-2619,
  and 2004,  
  September 7.4103  
         for IGR\,J16465-4507). Points in the central ($\pm0.5$
  day) part of the curves are given with better temporal
  resolution ($\sim 10^3$ s) and connected with a solid line (for clarity).}
\label{fig1}
\end{figure}

\noindent
sources of persistent X-ray emission $L_{\rm X}\sim
2.3\times10^{36}\ {P_*}^{-4/3}\ \dot{m}_*\ v_*^{-4} $
erg\,s$^{-1}$ due to accretion from the dense stellar wind (in
the case of circular orbits and without centrifugal inhibition
of accretion).  Here $\dot{m}_*=\dot{M}_w\,/10^{-5}\ M_{\odot}\
\mbox{yr}^{-1}$ and $v_* =v_{\rm w}/10^3$ km s$^{-1}$ are the
wind ejection rate and velocity, $P_*=P_b\,/10$ days. We assume
that mass and radius of the neutron star $M_1=1.4\ M_{\odot},$
$R_1=10$ km.

Fig.~\ref{fig1} shows the light curves from 6 SFXTs of the list
measured with the IBIS/ISGRI instrument on board INTEGRAL during
their first X-ray outbursts. Most of the curves demonstrate strong
flaring activity of the sources and have rather short details in
profile, although the curve of AX\,J1749.1-2733 is sufficiently
smooth. The intensity of sources rises by 4--5 orders of
magnitude during a very short time. Each outburst lasts for about
1 day in total. Such a duration is the main distinctive feature
of SFXTs. The only exception is IGR\,J11215-5952 with the
outbursts lasted for $\sim 5$ days, but this source may be
non-typical and flare up when its compact object approaches the
perigee \cite{sidoli07}. Other SFXTs have rather long intervals
of recurrence (months or years) significantly exceeding their
orbital periods. Many SFXTs do not show any regularity in the
outbursts at all.

The short life-time does not allow SFXTs to form an
accretion disk with a size comparable with the binary
separation. It is obvious that the outbursts are connected with
some type of instable accretion from the wind of an optical
star. However, accretion from the wind occurs on a free-fall timescale 
which is usually much shorter ($<10^3$ s) than the characteristic
life-time of SFXTs.

\noindent
The X-ray emission of SFXTs is characterized by a hard spectrum
extending to 15--200 keV. Being approximated by an optically
thin thermal bremsstrahlung it leads to the temperatures $kT\sim
10-30$ keV. The spectra of 6 SFXTs measured with IBIS/ISGRI
during their outbursts in 2003-2004 are shown in
Fig.\,\ref{fig2} together with their best-fit
approximation. Such spectra are typical for accreting neutron
stars. To illustrate this point we show in Fig.\,\ref{fig3}
the hard X-ray spectra of 2 bright SXBs, Cyg\,X-1 (which
harbours a black hole) and 4U\,1700-37 (harbours a neutron
star). We compare them with that of SAX\,J1818.6-1703. The
spectrum of Cyg\,X-1 is obviously the hardest one ($kT\sim75$
keV). The spectra of two other sources are much softer and
similar to each other if not identical.

\section{Models}
There were four models proposed to explain peculiar
properties of SFXTs and their outbursts:
\begin{itemize}\itemsep 0pt\parskip 1.3pt\topsep 0pt
\item the Be-type model assuming a very elliptical orbit for
  the binary, the outbursts are triggered at the moments when a
  compact object travels through its periastron
  \cite{chaty08,sidoli07}, 

\item the highly structured (clumpy) stellar wind from the
  supergiant, the outburst begins due to swallowing of one of
  the clumps of dense matter from the wind
  \cite{intzand04,negueruela08},

\item overcoming of a centrifugal barrier at the magnetospheric
   boundary of the neutron star which stops steady accretion
  onto its surface (the ``propeller effect''
  \cite{illarionov75}), the overcoming may occur due to even small
  increase in the wind local density or decrease in the wind velocity
  \cite{gs07},

\item overcoming of a magnetic barrier of the neutron star which
  could stop steady accretion onto its surface if the magnetic
  field of the neutron star $>10^{14}$ G and its spin
  period $>10^4$ s \cite{bozzo08}.
\end{itemize}
\noindent
The first effect is responsible for observed activity of
IGR\,J11215-5952 \cite{sidoli07}. Taking into account that the
$P_s$ and $P_b$ periods of AX\,J1749.1-2733 are close to those
of IGR\,J11215-5952 and that its light-curve differs
from the curves of other SFXTs in Fig.\,1, we can suggest that
this source also is a representative of Be-type SFXTs. For other
SFXTs the eccentricity $e$ seems to be too close to 1 to
explain the high $\sim100$ ratio between the recurrence
interval of outbursts and their duration.
\begin{figure}[p]

\vspace{-3mm}

\begin{minipage}[t]{0.6\textwidth}
\hspace{-8mm}\includegraphics[width=0.98\textwidth]{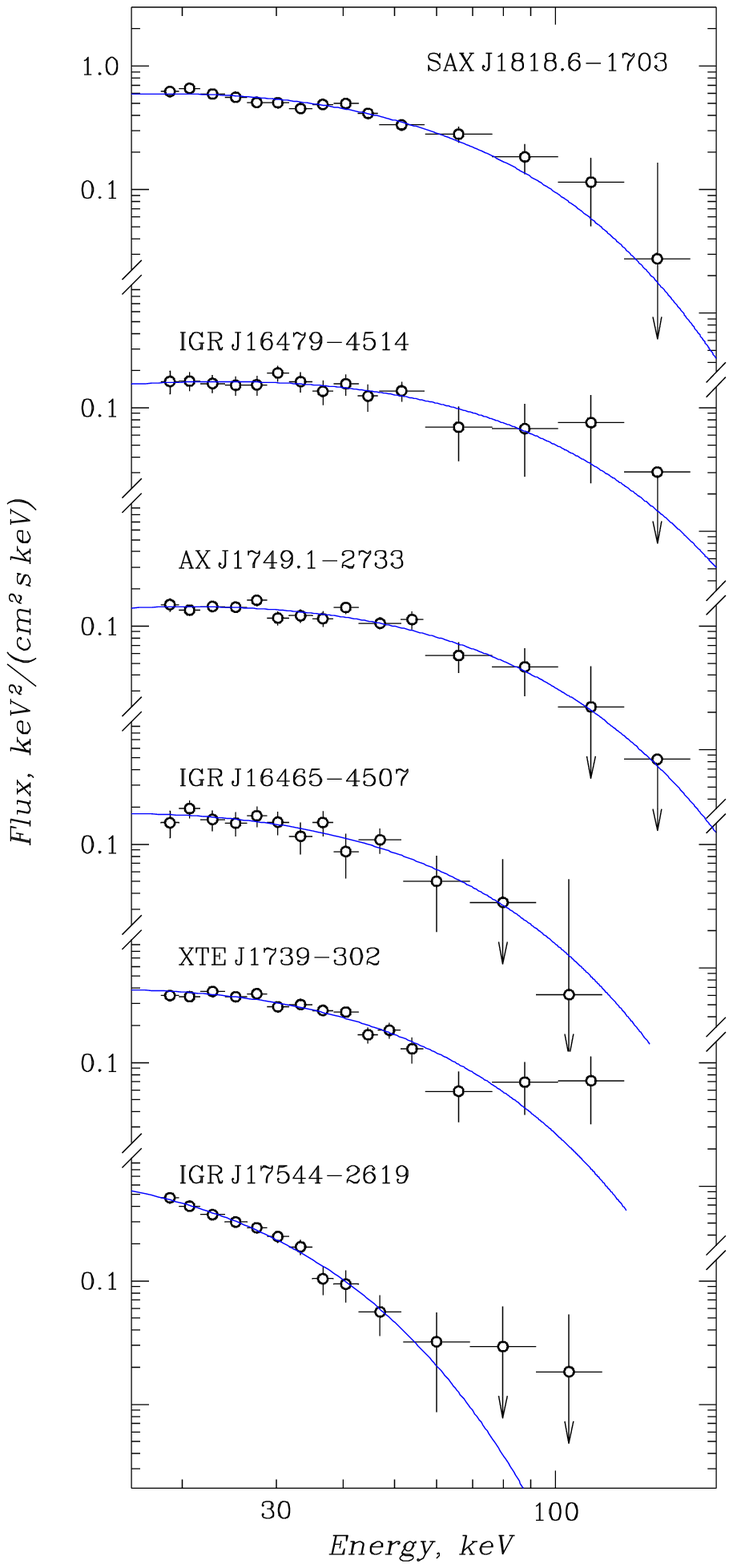}
\end{minipage}\hspace{-24mm}\begin{minipage}[b]{0.6\textwidth}
\hspace{5mm}\includegraphics[width=0.98\textwidth]{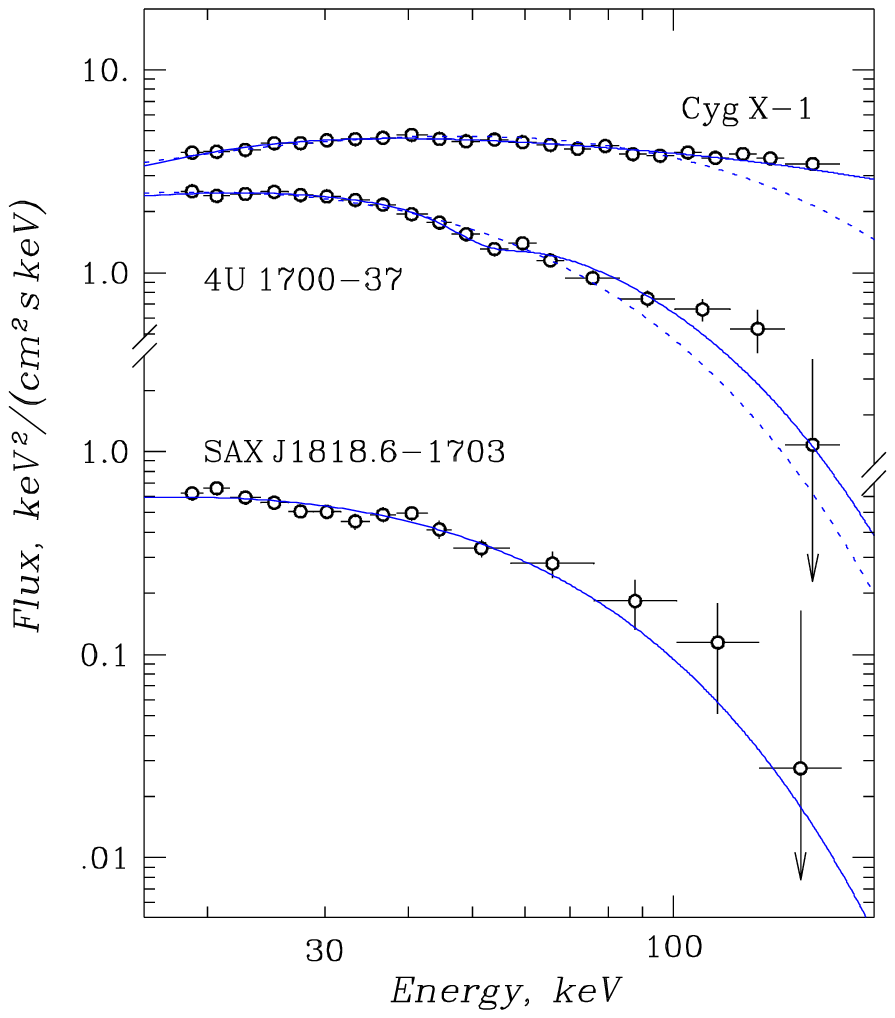}

\vspace{-10.55mm} 

\hspace{13.75mm}\includegraphics[width=0.85\textwidth]{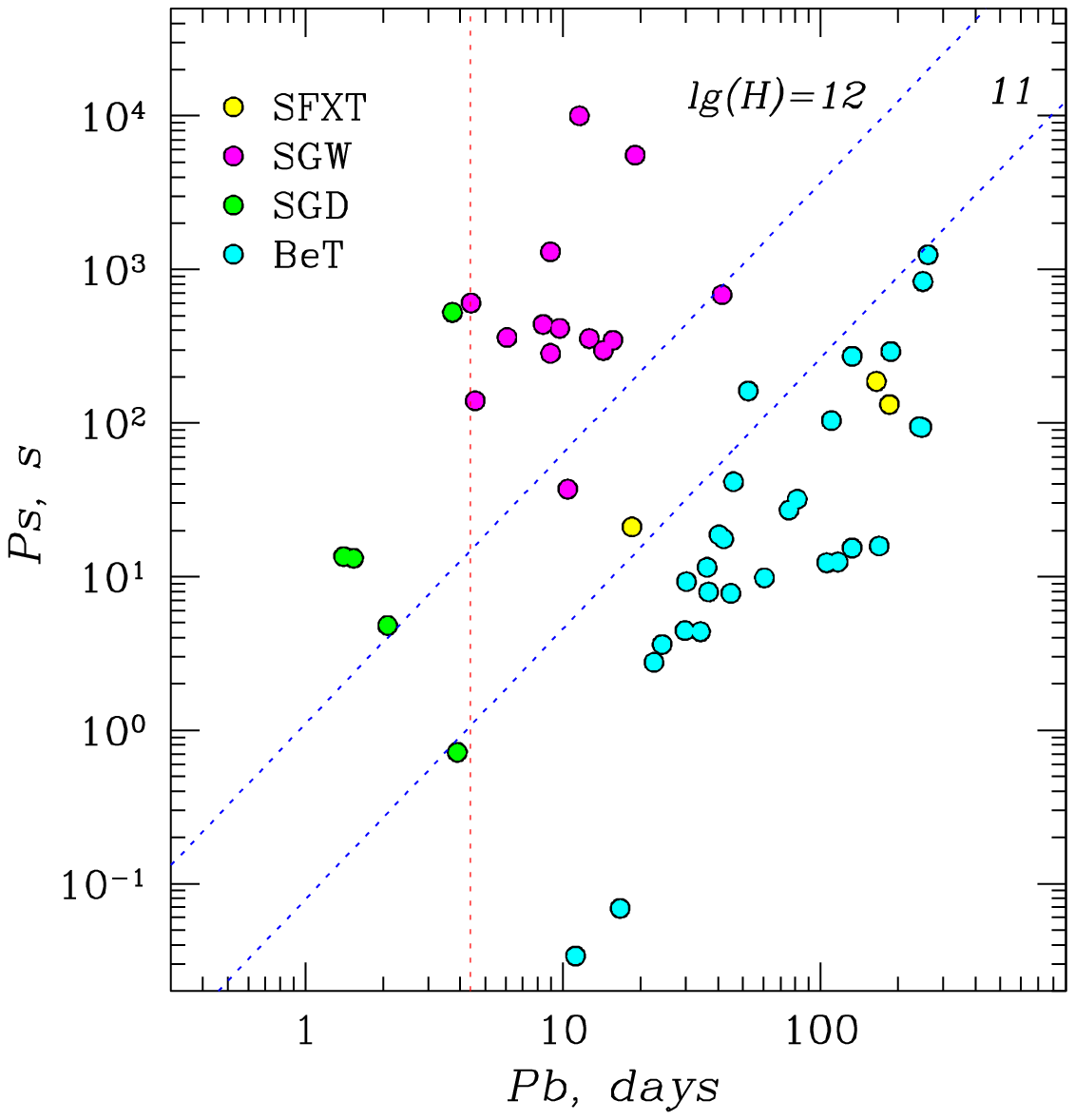}

\vspace{4.45mm} 
\end{minipage}

\vspace{-6mm} 
\caption{Hard X-ray spectra of 6 SFXTs (the same as in
  Fig.\,\protect\ref{fig1}) and their approximation with thermal
  bremsstrahlung. The spectra were obtained with IBIS/ISGRI in
  2003-2004 during the first detected outbursts of the
  sources. The best-fit temperature $kT=37.5\pm 6.3$ keV
  (IGR\,J16479-4514), $32.5\pm 3.1$ keV (AX\,J1749.1-2733),
  $28.8\pm 1.7$ keV (SAX\,J1818.6-1703), $24.0\pm 4.3$ keV
  (IGR\,J16465-4507), $22.4\pm 1.4$ keV (XTE\,J1739-302),
  $10.9\pm 0.8$ keV (IGR\,J17544-2619).}
\label{fig2}
\caption{IBIS/ISGRI spectra of two persistent SXBs, Cyg\,X-1
  (measured on June 8, 2003, during the intermediate state) and
  4U\,1700-37 (measured on September 12, 2003), in comparison
  with that of SAX\,J1818.6-1703. Dotted lines show the 
  approximation of these spectra with thermal bremsstrahlung,
  $kT=75.4\pm0.9$ keV for Cyg\,X-1, $30.6\pm0.5$ keV for
  4U\,1700-37 and $28.8\pm .7$ keV for SAX\,J1818.6-1703.}  
\label{fig3}
\caption{$P_s-P_b$ diagram for HMXBs. Different colours show
  SFXTs, disk-fed (SGD) and wind-fed (SGW) supergiant binaries,
  and Be-transients (BeT). The red line indicates $P_b$ at which
  the supergiant with $R=20 R_{\odot}$ and $M=22 M_{\odot}$
  fills its Roche lobe, blue lines -- $P_s$ below which the
  centrifugal barrier inhibits quasi-spherical accretion
  from the stellar wind (for $H=10^{11}$ and $10^{12}$ G,
  $v_w=800\ \mbox{km s}^{-1}$).}
\label{fig4}
\end{figure}

In the clumpy wind model the contrast $\xi$ of the gas density
in clumps relatively to the average density should be of $\sim
10^{4}$ to explain the X-ray luminosity ratio between outburst
and quiescent states of SFXTs. It is difficult to believe that
such dense clumps can exist in the hot wind of OB
supergiants. But even if they exist the compact object should
intersect and swallow them on a time scale of
$a\,\xi^{-1/3}v_{\rm w}^{-1}\sim10^3$ s that is much shorter
than the life-time of SFXTs. Here $a$ is the binary
separation, $a=[G(M_1+M_2)]^{1/3} (P_b/2\pi)^{2/3}\simeq 62\
P_*\ R_{\odot}$, and $M_2\simeq30\ M_{\odot}$ is the companion
mass.

Two latter models of the list can explain the
outburst/quiescence luminosity
ratio with ease. They assume that a compact object in SFXTs is a
neutron star with sufficiently strong magnetic field $H=10^{12}
h_*$ G. We have already shown that this is really the case. The
magnetic inhibition regime takes place when the accretion radius
of the neutron star $R_a=2GM_1/v_{\rm w}^{\,2}\simeq 0.54\ v_*^{-2}\
R_{\odot}$ is smaller than its magnetospheric radius $R_m=[0.5
H^{\,2} R_1^{\,\,6} /\dot{M}/(2GM_1)^{1/2}]^{2/7}\simeq 0.0066\ h_*^{\,4/7}
v_*^{\,8/7} P_*^{\,\,8/21} \dot{m}_*^{-2/7}\ R_{\odot}$ where
$\dot{M}=0.25 (R_a/a)^2 \dot{M}_w\simeq 1.9\times10^{-10}
v_*^{\,-4} P_*^{\,-4/3} \dot{m}_*\ M_{\odot} \mbox{yr}^{-1}$ is the
accretion rate. This regime works when $H$ is in the magnetar
range $>10^{14}$ G that is unlikely for SFXTs \cite{bozzo08}.

The propeller regime takes place when $R_m$ exceeds the
corotation radius of the neutron star $R_c=(GM_1/\Omega^2)^{1/3}$
$\simeq 0.0024\ P_s^{\, 2/3}\ R_{\odot}$ remaining smaller than
$R_a$ ($R_{c}<R_{m}<R_{a}$). The magnetospheric radius depends
on accretion rate (or the stellar wind local density and
velocity), so if $R_{c}$ differs only slightly from $R_{m}$
transitions between propeller and direct accretion regimes may
occur even due to rather a small increase in the wind density
(or decrease in its velocity). The equilibrium $R_{c}=R_{m}$
takes place when the neutron star's spin period $P_s^{\,\,*}\simeq
4.5\ h_*^{6/7} \dot{m}_*^{-3/7} v_*^{\,12/7} P_*^{\,\,4/7}\, \mbox{s}$.
According to \cite{gs07} SFXTs should have $P_s$ slightly
smaller than $P_s^{\,\,*}$. They become bright in X-rays when the wind
density increases in $(P_s^{\,\,*}/P_s)^{7/3}$ times and $R_m$ drops
below $R_c$.  The outburst duration is determined by the time
that the neutron star spent in the region of enhanced
density. If its orbit is circular the enhancement may be related
to only rather a long change in the wind rate. It should be an
intrinsic property of the supergiant amplified by the effect of its
surface heating by X-rays from the outburst.
  
Using $R_c=(GM_1/\Omega^2)^{1/3}$ in such estimates in
\cite{gs07} we have compared the magnetospheric velocity with
the Keplerian one $v_{K}=(GM_1/R_c)^{1/2}$. In reality the specific
angular momentum that accreting matter carries to the neutron star
$j=\pi R_a^{\,2}/P_b$ is smaller than the Keplerian value
\cite{illarionov75,bkogan91}. The velocity at the corotation
radius should be $v=\pi R_a^{\,2}/R_c/P_b$, the radius itself
$R_c =\!(0.5 P_s/P_b)^{1/2}R_a$ $\simeq0.00041\
P_{\,\,*}^{\,\,-1/2}P_s^{\,\,1/2}v_*^{\,-2}\ R_{\odot}$, and the equilibrium period
$P_s^{\,\,*}\simeq 258\ P_*^{\,\,37/21} h_*^{\,8/7} v_*^{\,44/7}
\dot{m}_*^{-4/7}\ \mbox{s}$.

In Fig.\,\ref{fig4} we show by blue lines this dependence for
two reliable values of the neutron star's magnetic field 
$H=10^{11}$ and $10^{12}$ G. 
The figure shows the $P_s-P_b$ diagram for a large sample of
HMXBs. Three distinct populations of sources are seen: wind-fed
and disk-fed SXBs and Be-transients. We expect SFXTs to appear
in the diagram somewhere between the blue lines. Unfortunately
of 3 SFXTs with currently known $P_s$ and $P_b$ \ 2 already
mentioned sources are located among Be-transients, and only the
last one, IGR\,J18483-0311, approved our expectations. Further
search for all types of periodicities in\,SFXTs is\,extremely
important for understanding of\,their nature.

\acknowledgments This research was supported by the program
``The origin, structure and evolution of objects of the
Universe'' of the Presidium of RAS and the grant NSh-5069.2010.2
of the Russian President.


\begin{thebibliography}{99}\parskip 1.9pt\topsep 0pt

\bibitem
{chaty08} Chaty S., 2008,
\emph{Chin. J. Astron. Astrophys. Suppl.\/}, {\bf 8}, 197


\bibitem
{bkogan91} Bisnovatyi-Kogan G.S., 1991, \emph{A\&A\/}, {\bf 245}, 528

\bibitem
{bozzo08} Bozzo E., Falanga M., Stella L., 2008, \emph{ApJ\/},
{\bf 683}, 1031 



\bibitem
{gs05} Grebenev S.A. \& Sunyaev R.A., 2005, \emph{AstL\/}, {\bf 31}, 672

\bibitem
{gs07} Grebenev S.A. \& Sunyaev R.A., 2007, \emph{AstL\/}, {\bf 33}, 149



\bibitem
{illarionov75} Illarionov A.F. \& Sunyaev R.A., 1975, \emph{A\&A\/},
{\bf 39}, 185


\bibitem
{intzand04} in't Zand J., Heise J., Ubertini P., Bazzano A.,
Markwardt C., 2004, 
\emph{ESA SP\/}--{\bf 552}, 427  






\bibitem
{negueruela06} Negueruela I., Smith D.M., Reig P., Chaty S.,
Torrejon J.M., 2006, 
\emph{ESA SP\/}--{\bf 604}, 165


\bibitem
{negueruela08} Negueruela I., Torrej\'on, J. M., Reig, P.,
Rib\'o, M., Smith, D.M., 2008, 
\emph{AIPC\/}, {\bf 1010}, 252

\bibitem
{sidoli07} Sidoli L., Romano P., Mereghetti S., Paizis
A., Vercellone S., et al.,
2007, \emph{A\&A\/}, {\bf 476}, 1307

\bibitem
{sguera06} Sguera V., Bazzano A., Bird A.J., Dean A.J.,
Ubertini, P., et al.
, 2006, \emph{ApJ\/}, {\bf 646}, 452 


\bibitem
{smith04} Smith D.M., 2004, \emph{ATel\/}, {\bf 338}


\bibitem
{sunyaev03b} Sunyaev R.A., Grebenev S.A., Lutovinov A.A.,
Rodriguez J., Mereghetti S., et al.,
2003, \emph{ATel\/}, {\bf 190}

\bibitem
{cw03} Winkler C., Courvoisier T.J.-L., Di Cocco G., Gehrels N.,
Giménez A., et~al., 2003, \emph{A\&A\/}, {\bf 411}, L1
\vspace{-3mm}

\end{thebibliography}
\end{document}